\documentclass[11pt,twoside]{article}
\usepackage{./asp2014}

\aspSuppressVolSlug
\resetcounters

\begin{document}

\title{Polarimetry as a tool to study multi-dimensional winds and disks}
\author{Jorick S. Vink$^1$ 
\affil{$^1$Armagh Observatory, College Hill, BT61 9DG Armagh, Norn Iron; \email{jsv@arm.ac.uk}}}


\begin{abstract}
I start with a discussion of spherical winds and small-scale clumping, before continuing with various theories that 
have been proposed to predict how mass loss depends on stellar rotation -- both in terms of wind strength, as well as the 
latitudinal dependence of the wind. This very issue is crucial for our general  
understanding of angular momentum evolution in massive stars, and the B[e] phenomenon in particular. 
I then discuss the tool of linear 
polarimetry that allows us to probe the difference between polar and equatorial mass loss, allowing us to test B[e] and related 
disk formation theories. 
\end{abstract}

\section{Introduction: what makes B[e] stars special?}

The role of B[e] stars, and the B[e] supergiants in particular, in stellar evolution 
is still very much open. One outstanding aspect is that 
of their rapid rotation rates (see Kraus these proceedings) at significant
levels of the stellar brake-up speed. 
It is this unique aspect that may provide clues
to their origin, as the general population of canonical B supergiants are {\it slow}
rotators (see Vink et al. 2010 and Fig.\,\ref{vrot}).
The unique rapid rotation of B[e] stars (and also some of the S Doradus-type 
luminous blue variables (LBVs; Groh et al. 2006) may hint at a merger origin (see Vanbeveren these proceedings; 
Podsiadlowski et al. 2006; Justham et al. 2014).

Whether B[e] supergiants are ultimately related to single star or binary star evolution
is one question, but the other aspect concerns the way mass is lost from 
their surfaces, and how this affects angular momentum evolution in 
massive stars towards explosion (see Georgy, these proceedings). 
In order to test mass-loss predictions for rotating massive stars, we 
need to find a method that can probe the density contrast between the 
stellar pole and equator. 

In the local Universe, this may be achieved through 
long-baseline interferometry, as discussed during this meeting by Meilland, but 
in order to determine wind asymmetry in the more 
distant Universe we need to rely on the technique of {\it linear} 
spectropolarimetry. 
In Sect.\,2 we first discuss spherical radiation-driven winds, including wind clumping and 
the relatively new aspect of the Eddington $\Gamma$ parameter in mass-loss predictions,  before turning to 2D models in Sect.\,3. 
The second part of this review involves observational tests with linear $QU$ line 
spectropolarimetry. 

\articlefigure[width=.7\textwidth]{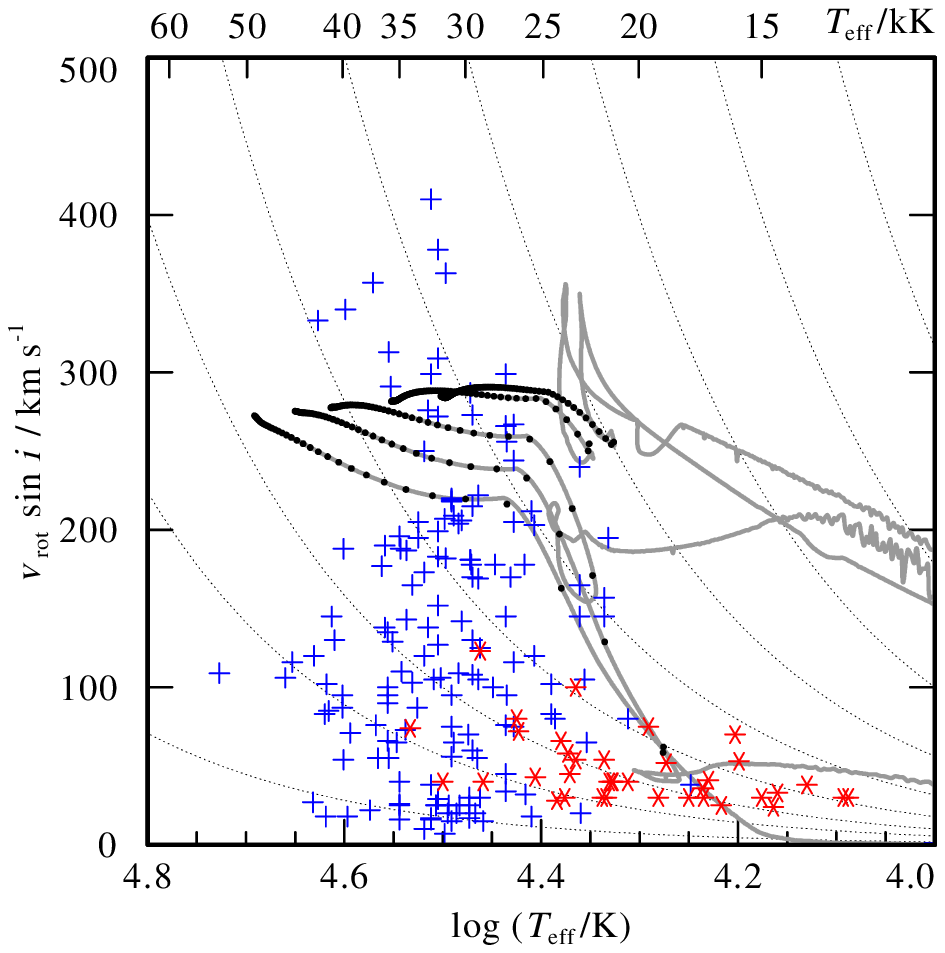}{vrot}{Rotational velocities as a function of 
$T_{\rm eff}$ for OB stars from the VLT-Flames survey of massive stars  
with evolutionary masses above 15\,$M_\odot$. 
Luminosity classes are shown as blue pluses (luminosity classes {\sc ii-v}) 
and red stars (luminosity class {\sc i}). The LMC evolutionary tracks from Brott et al. (2011), which include 
the bi-stability-jump, are 
shown in grey with initial $\varv_{\rm rot}=250$\,km/s for five masses of 15, 20, 30, 40 and 60\,$M_\odot$. 
It can be seen that the critical mass for bi-stability braking in the LMC is $\sim$35$M_{\odot}$.
The steepness of these tracks may be compared to the case of angular momentum conservation (drawn 
as grey dotted background lines).
The black dots on the tracks represent time-steps of $10^5$ years. 
See Vink et al. (2010) for more details.}

\section{Spherical winds: from CAK to a new formulation}

For optically thin O-star winds two distinct approaches
are in use to compute the relevant (line) opacity $\kappa$
and the resulting radiative acceleration $g_{\rm rad}$.
In the first method, the
line acceleration is expressed as a function of the velocity gradient $(dv/dr)$
in the CAK theory of Castor et al. (1975). 
In the second approach, the line acceleration is expressed as a function of
radius $r$ (Lucy \& Solomon 1970), which
has been implemented in Monte Carlo models, see Fig.\,\ref{gline} (M\"uller \& Vink 2008).

\articlefigure[width=.9\textwidth]{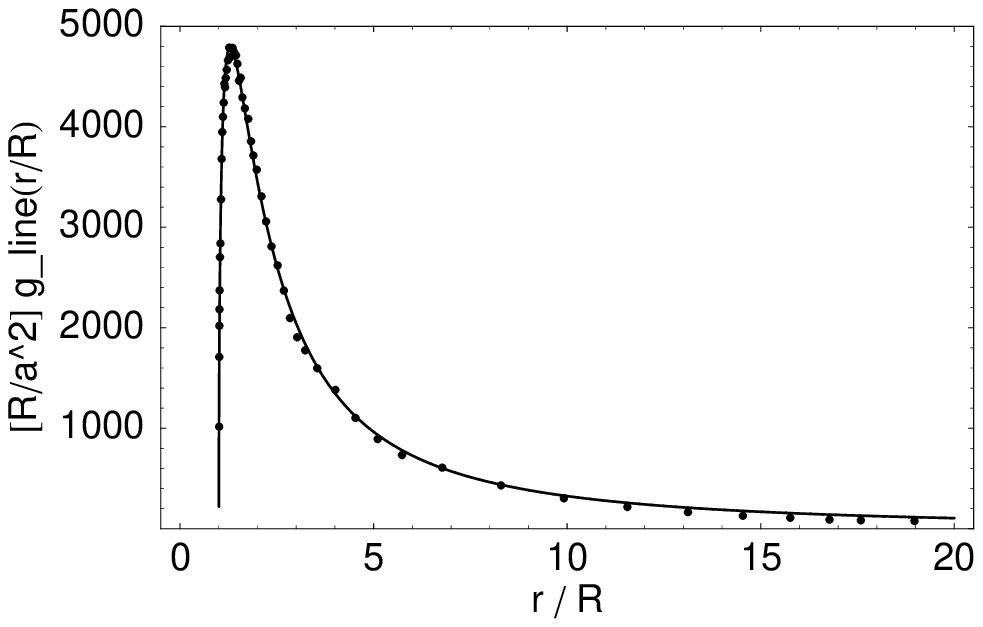}{gline}{The dots represent the dimensionless radiative line acceleration 
         ${g_{\rm rad}^{\rm line}}\,({r})$ 
         versus radial distance $r$ for a typical O5V-star, as derived from Monte Carlo simulations.
	 In order to determine the line acceleration parameters, 
	 the simulated values were fitted with a non-linear model equation (see M\"uller \& Vink 2008) resulting in 
         the solid curve (wind parameters then follow from an iteration process).}

\subsection{Instabilities and wind clumping}

Because of the non-linear character of the equation of motion, the CAK solution is complex, with the
physics involving instabilities due to the line-deshadowing instability LDI (see Owocki 2015). 
One of the key implications of the LDI
is that the time-averaged $\dot{M}$ is {\it not}
expected to be affected by wind clumping, as
it has the same (average) $\dot{M}$
as the CAK solution. However, the shocked velocity structure and 
associated density are expected to result in changes of the mass-loss
diagnostics. 

$\dot{M}$ diagnostics that is dependent
on the square of the density, such as H$\alpha$, will
result in a square-root reduction of $\dot{M}$ (Hillier 1991; Hamann et al. 2008), whilst 
ultraviolet P Cygni lines such as 
P{\sc v} depend linearly on 
density (Puls et al. 2008). 
Most analyses have been based on the assumption of optically thin clumps, but 
clumped winds are porous, with a range of clump
sizes and optical depths. 
Oskinova et al. (2007) 
employed an effective opacity concept 
for line-profile modelling of the O supergiant $\zeta$ Pup, showing 
that the most pronounced 
effect involves resonance lines like P{\sc v}, which can be
reproduced by their macro-clumping approach (see also Sundqvist et al. 2010).

In the traditional view of line-driven winds of O-type stars via the 
CAK theory and the associated LDI, clumping would be expected to develop
in the wind when the wind velocities are large enough to produce shocked
structures. For O star winds, this is thought to occur at 
about half the terminal wind velocity at about 1.5 stellar radii.
Observational indications from linear 
polarization (e.g. Davies et al. 2005; 2007) however show that clumping already 
exists close to the stellar photosphere.
Cantiello et al. (2009) suggested that convection in the sub-surface 
layers associated with the iron opacity peak could be the root 
cause of wind clumping. 

\subsection{High Eddington $\Gamma$ factor}

We have discussed the optically thin stellar winds of normal O stars, but 
at certain luminosities the winds may become optically thick. What this means is that the wind optical depth 
$\tau$ crosses unity and thereby also the wind efficiency number $\eta$
crosses unity (Vink \& Gr\"afener 2012, using data and model results by Martins et al. 2008). 

For (super)O-stars (on steroids) this means the absorption line dominated
spectrum turns into a Wolf-Rayet type spectrum of the hydrogen-rich (and nitrogen (N) 
rich) variety: WNh. 
Bestenlehner et al. (2014) studied this transition
between optically thin (O star) and optically thick (Of/WN; WNh) stars
in the context of the VLT-Flames Tarantula survey (VFTS; Evans et al. 2011) 
by analyzing 62 objects with {\sc cmfgen} (Hillier \& Miller 1998), 
confirming the predicted kink of Vink et al. (2011), see Fig.\,\ref{gamma}.

\articlefigure[width=.9\textwidth]{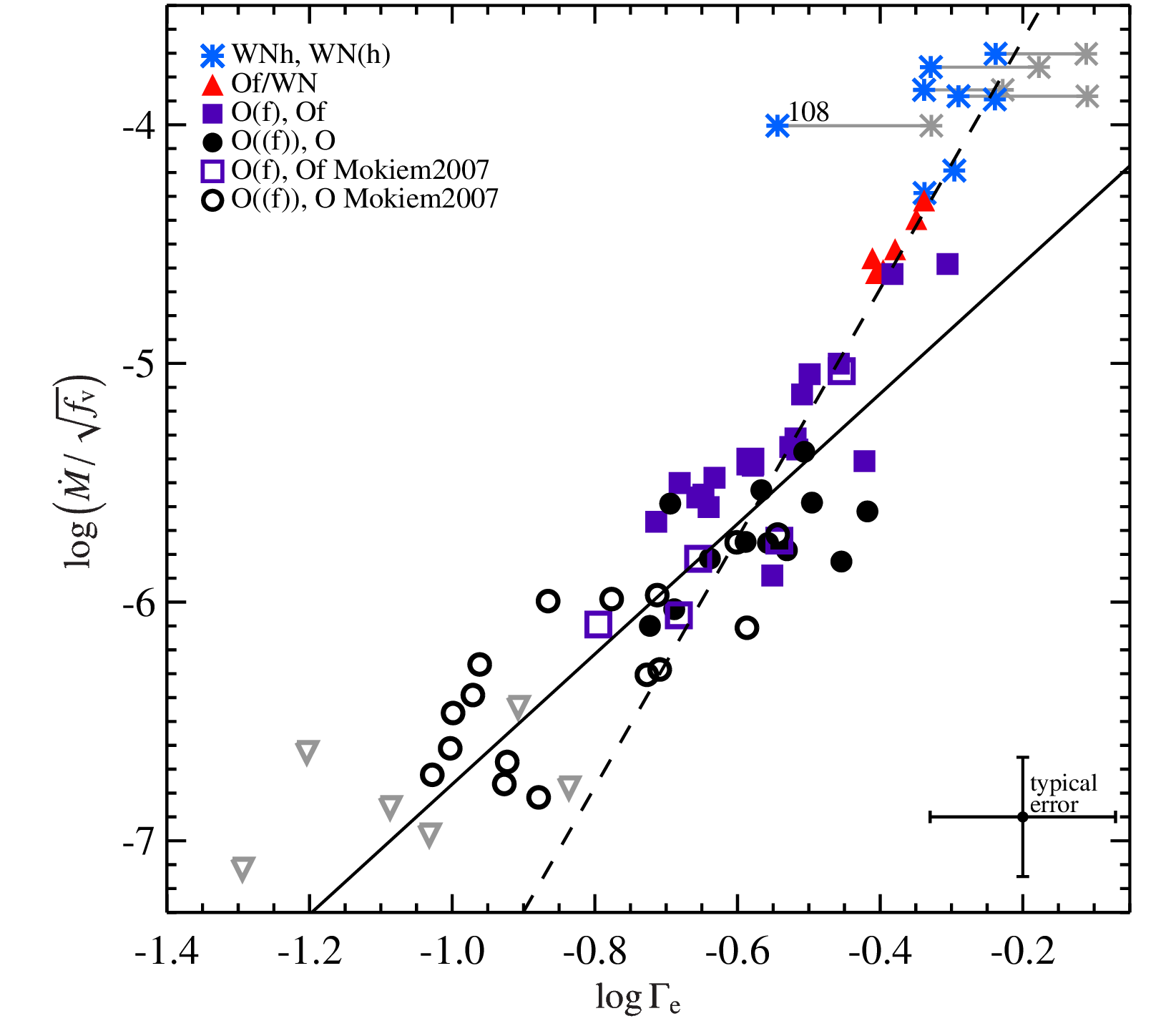}{gamma}{Unclumped 
$\log \dot{M}$ vs.~$\log \Gamma_{\rm e}$. Solid line: $\dot{M}-\Gamma_{\rm}$ relation for O stars. The
  different symbols indicate stellar sub-classes. Dashed line: the steeper slope
  of the Of/WN and WNh stars -- forming a {\it kink}. 
   The grey asterisks indicate the
    position of the stars with $Y > 0.75$ under the assumption of
    core He-burning. The grey upside down triangles are stars from
  Mokiem et al. (2007) which only have an upper limit in $\dot{M}$ 
are excluded from the fit. See Bestenlehner et al. (2014).}

This means that mass-loss rates in stellar evolution should not be considered 
as a function of luminosity, but the $L/M$ ratio related to the Eddington $\Gamma$ factor instead (Vink 2015). 
Another relevant factor (besides metallicity) is that of the effective temperature (see below).

\subsection{The Bi-stability jump}

At certain specific effective temperatures, 21\,000 K, and again at 10\,000 K
the ionization of iron (Fe) changes dramatically from Fe {\sc iv} to Fe {\sc iii}, and at cooler 
temperatures from Fe {\sc iii} to Fe {\sc ii}, causing a dramatic increase in the line acceleration
below the sonic point and in the predicted mass-loss rate as a result (Vink et al. 1999; Petrov et al. 2016). 
These transitions are referred to as bi-stability jumps (Pauldrach \& Puls 1990; Lamers et al. (1995).
These jumps may play an important role in stellar wind braking as discussed above, but the mechanism
may also play a role in the 2D latitudinal dependence of winds between a hotter pole and a cooler
equator, as in the rotationally-induced bi-stability mechanism of Lamers \& Pauldrach (1991) for B[e] stars 
that is shown in Fig.\,\ref{rib} and discussed in more detail below. 

\section{2D stellar winds and disks}

Until 3D radiation transfer models with 3D 
hydrodynamics become available, theorists 
have been forced to make assumptions with respect
to either the radiative transfer (e.g. by assuming a power law
approximation for the line force due to CAK
or the hydrodynamics, e.g. by assuming an empirically motivated 
wind terminal velocity in Monte Carlo predictions (Abbott \& Lucy 1985). 
Although recent 1D and 2D models of M\"uller \& Vink (2008; 2014) no longer 
require the assumption of an empirical terminal wind velocity. 

There are 2D wind models on the market that 
predict the wind mass loss predominately emanating
from the equator, whilst other models predict 
higher mass-loss rates from the pole. 
The first effect of stellar rotation that needs to be taken into account is that 
of a reduced effective gravity at the stellar equator (Friend \& Abbott 1986), with 
2D modifications and expansions by e.g. Bjorkman \& Cassinelli (1993) and Cure (2004).
These 2D equatorial wind models might potentially result in the required equatorial disk 
formation for Be and B[e] stars. However, due to a combination of non-radial line forces 
and the result of the Von Zeipel (1924) effect -- resulting 
in a hotter pole -- equatorial disk formation is suppressed, and a stronger 
polar wind is expected instead (Owocki et al. 1996, Petrenz \& Puls 2000). 
Furthermore, disks might be ablated due to the strong radiation fields around 
OB stars (Kee et al. 2016).

Is there no hope to form an equatorial outflow using radiative pressure?
The best physical model that is available is that of the 
rotationally-induced bi-stability model of Lamers \& Pauldrach (1991), 
improved with appropriate Fe line driving models in Pelupessy et al. (2000).
The key point of this model is that there is a natural temperature 
transition as a function of stellar latitude as can be noted in Fig.\,\ref{rib}.
The hot pole drives a fast (several 1000s km/s) hot-star wind driven by
high ionization stages (since Vink et al. (1999) we know this should 
be Fe {\sc iv} and not C {\sc iv}), whilst the cooler equator drives a slow wind 
(100s of km/s) driven by lower ionization stages of Fe.

\articlefigure[width=.9\textwidth]{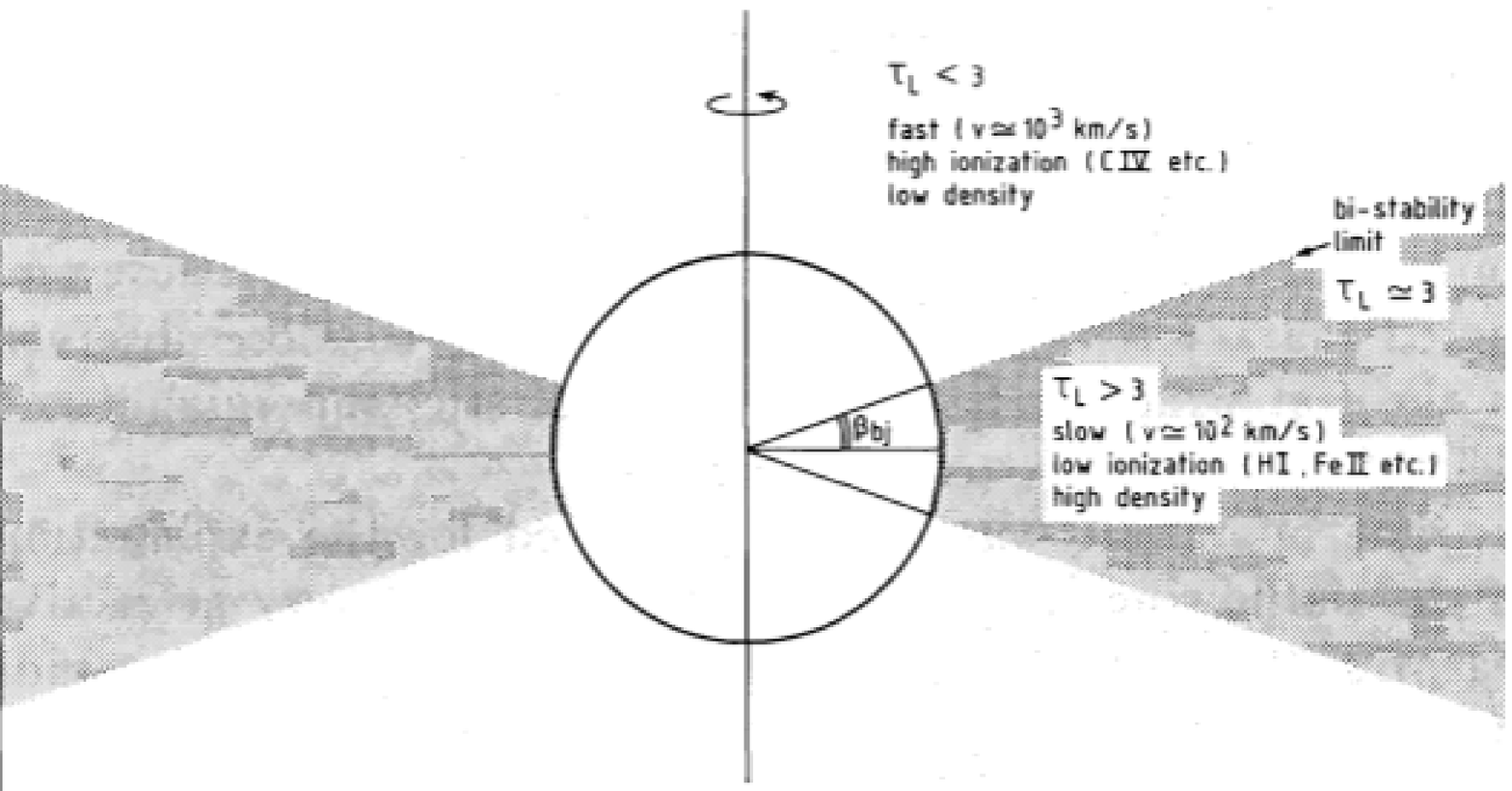}{rib}{Note the temperature 
transition as a function of stellar latitude due to Von Zeipel (1924).
The hot pole drives a fast (several 1000s km/s) hot-star wind driven by
high ionization stages of Fe {\sc iv}, whilst the cooler equator drives a slower 
wind (100s of km/s) driven by lower ionization stages of Fe. 
Figure from Lamers \& Pauldrach 1991.}

The author is well aware that for Be stars, and now also B[e] stars (see later), the 
disks seem to be in Keplerian motion, with outflow velocities being
less evident. Such rotating disks may well be described by a viscous disk (see 
Lee et al. 1991; Okazaki 2001; Carciofi et al. 2012) but such a viscous disk would still not 
explain {\it why} disk formation occurs in the first place.
The prevalence of disks at spectral types B (in close proximity to the 
temperature of the bi-stability jump) means that the rotationally-induced 
mechanism remains an attractive one.

In any case, the key point from the different theoretical options is 
that mass loss from the equator results in more angular 
momentum loss than would 1D spherical or 2D polar mass loss, so we need 
2D data to test theoretical models.

\section{The tool of spectropolarimetry}

The principle of linear spectropolarimetry is quite simple. The tool
is based on the assumption that free electrons in an extended 
circumstellar medium scatter continuum radiation from the central star,  
revealing a certain level of linear polarization. 
If the projected electron distribution is perfectly 
circular, e.g. when the gas is spherically symmetric 
or when a disk is observed face-on, the linear 
Stokes vectors $Q$ and $U$ cancel, and no polarization 
is detected (as long as the object is spatially unresolved). 
If the geometry is not circular but involves  
an inclined disk, this would result in net continuum polarization.  

\articlefiguretwo{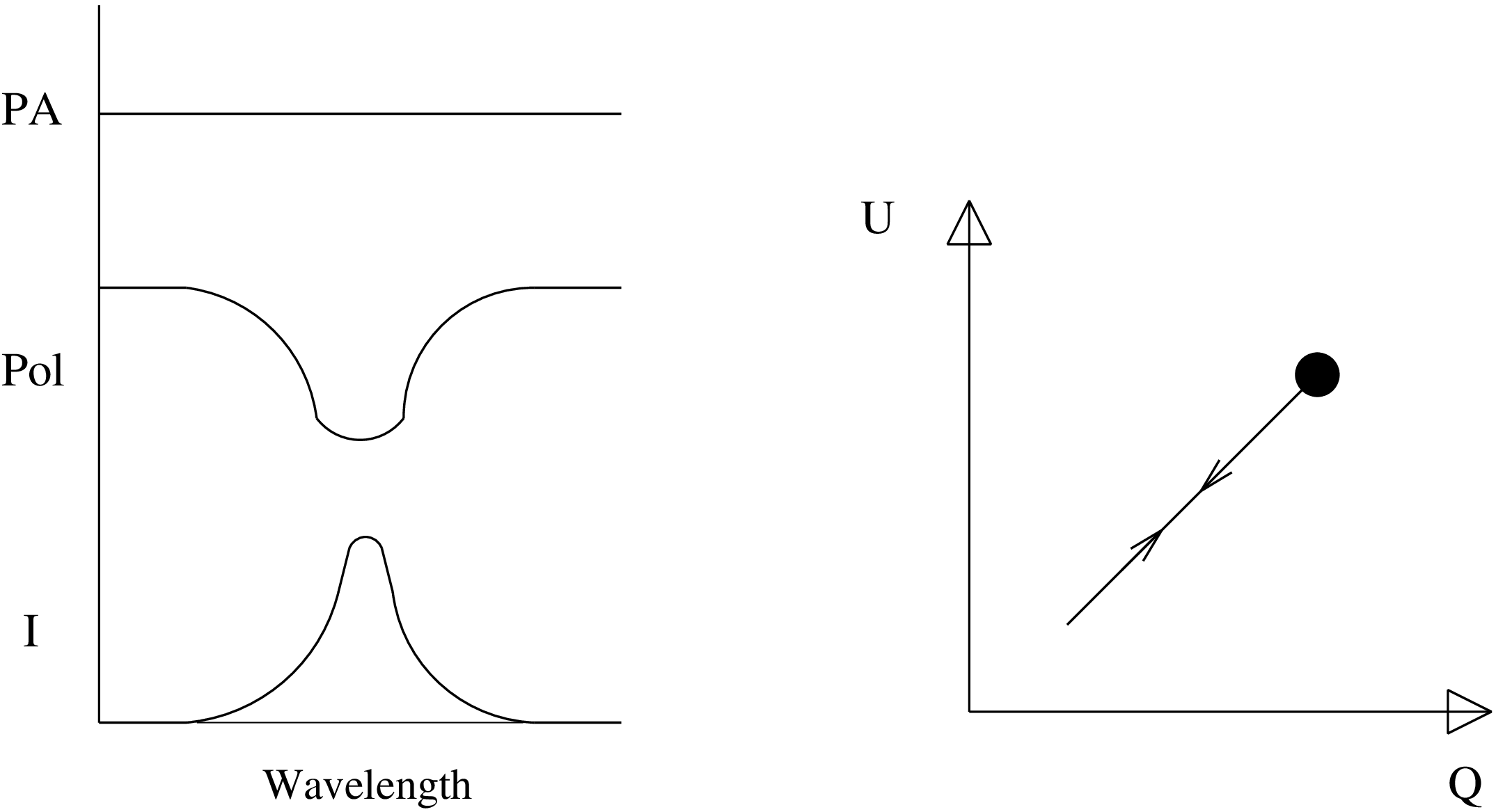}{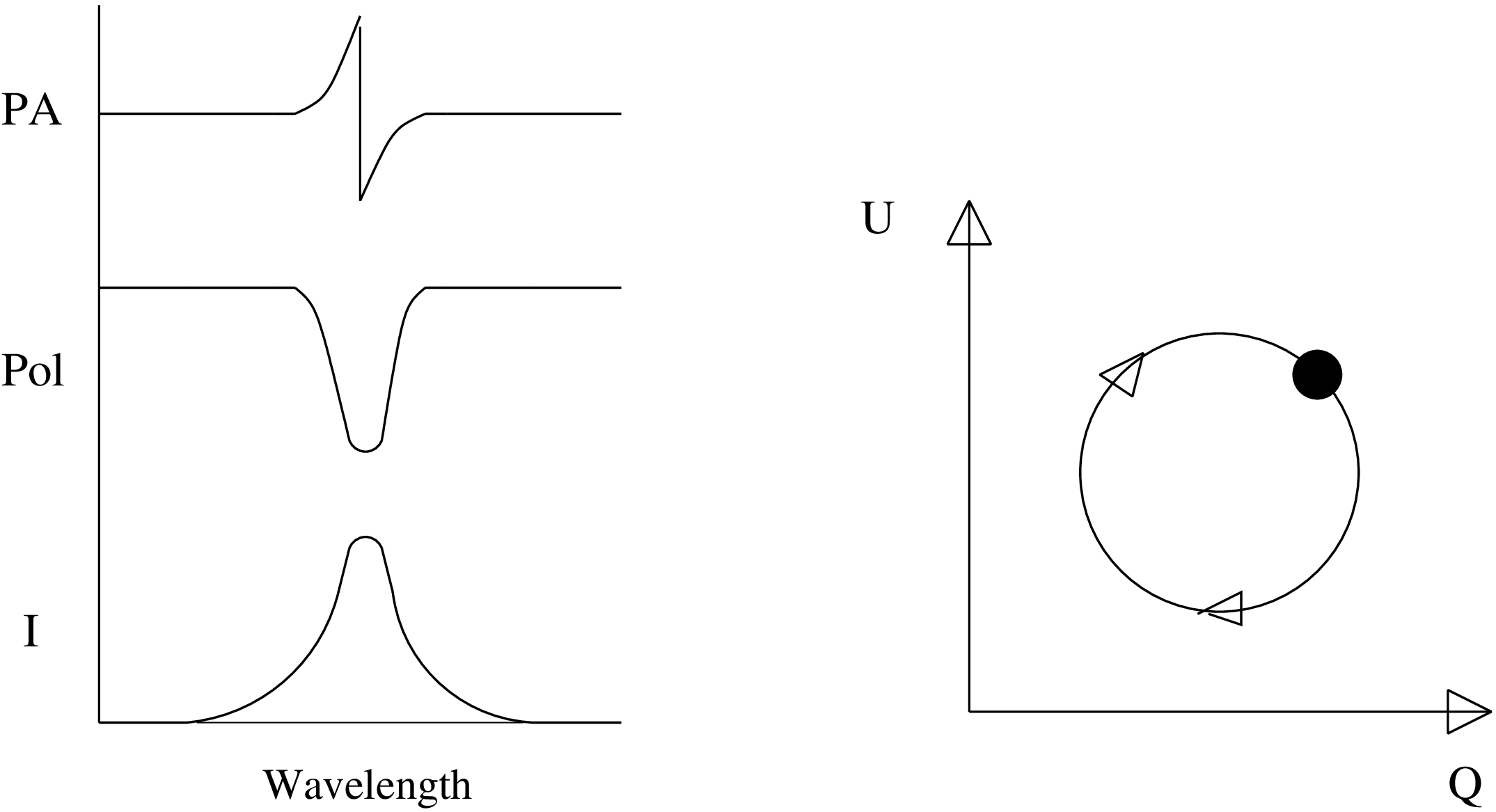}{cartoons}{Cartoons representing 
line {\it depolarization} (left hand side) and compact line emission
scattered off a rotating disk (right hand side) as triplots and $QU$ diagrams. Stokes $I$
profiles are shown in the lower triplot panels, \% Pol in the middle panels, and 
the position angles (PAs) are given in the upper triplot panels. 
Line depolarization is as broad as the Stokes I emission, while the {\it line} 
polarization is narrow by comparison. Depolarization translates 
into $QU$ space as a linear excursion (left hand side), 
whilst a {\it line} polarization PA flip is associated with a $QU$ loop (right hand side).}

One of the advantages of spectropolarimetry over continuum polarimetry is 
that one can perform differential polarimetry between a spectral line and the 
continuum {\it independent} of instrumental/interstellar 
polarization (ISP). The H$\alpha$ depolarization ``line effect'' utilizes the expectation 
that hydrogen recombination lines arise over a much larger volume than the 
continuum and becomes {\it de}polarized (see the left hand side of 
Fig.\,\ref{cartoons}). 
Depolarization immediately indicates the presence (or absence) of 
aspherical geometries, such as disks, on spatial scales that cannot be 
imaged with the world's largest telescopes. 

The basic idea of the technique was explored in the 1970s by e.g. Poeckert
\& Marlborough (1976) who employed narrow-band filters to show that
Be stars have disks as around 55\% of their objects showed the 
depolarization line effect. 
It took another couple of decades before interferometry 
confirmed these findings. Interestingly, in a recent 
study of peculiar O stars, Vink et al. (2009) 
did not find evidence for the presence of disks in Oe stars -- the alleged 
counterparts of classical Be stars -- although the first detection of a line effect 
in an Oe star (HD\,45314) was also reported. 

In general we divide the polarimetric data into bins 
corresponding to 0.1\% polarization, the typical error bar (although 
the numbers from photon statistics are at least a factor 10 better). 
We also present the data in $QU$ diagrams. 
For the case of line depolarization this translates into 
a linear excursion from the clusters of points that represents the continuum 
($P^2 = Q^2 + U^2$) with the excursion showing the trend 
when the polarization moves in and out of line center.  

A most relevant quantity involves the detection limit, which is inversely 
dependent on the signal-to-noise ratio (SNR) and the contrast of the emission line
to the continuum. The detection limit $\Delta P_{\rm limit}$ can be represented by:

\begin{displaymath}
\Delta P_{\rm limit} (\%) = \frac{100}{SNR} \times \frac{l/c}{l/c-1}
\end{displaymath}

\noindent where $l/c$ refers to the line-to-continuum contrast. 
This detection limit is most useful for objects with strong emission lines, such 
as H$\alpha$ emission in Luminous Blue Variables (LBVs; Humphreys \& Davidson 1994; Smith et al. 2004; Vink 2012), 
where the emission completely overwhelms underlying photospheric absorption. 
In general, one wishes to achieve a signal-to-noise ratio (in the continuum) of 
about 1000, corresponding
to changes in the amount of linear polarization of 0.1\%. 
We should be able to infer asymmetry degrees in the form of 
equator/pole density ratios, $\rho_{\rm eq}/\rho_{\rm pole}$ of $\sim$1.25, or larger 
(Harries 2000), with some small additional dependence on the shape and 
inclination of the disk.

Most of the linear line polarimetry work of the last two decades has indeed concerned
line depolarization, but in the following we will see that in some cases there is 
evidence for intrinsic {\it line} polarization, predicted by Wood et al. (1993) and 
found observationally in pre-main sequence (PMS) T Tauri and Herbig Ae/Be stars by our group 
(Vink et al. 2002, 2003, 2005b, Mottram et al. 2007).
In such cases line photons are thought to originate from a {\it compact} source, e.g.
as a result of (magnetospheric) accretion. These compact photons are scattered off a rotating 
disk, leading to a flip in the position angle (PA), and resulting in a rounded loop (rather than a linear excursion) 
in the $QU$ diagram (sketched on the right hand side of Fig.\,\ref{cartoons}).

\section{Depolarization results for B[e] supergiants and LBVs}

Line depolarization has been observed in a plethora of massive stars, involving 
B[e] supergiants (e.g. Oudmaijer \& Drew 1999), post Red 
Supergiants (Patel et al. 2008), as well as LBVs (see below). In all these 
cases the incidence rate of ``line effects'' appears to be consistent with the Be star 
results, i.e. 50-60\%. Furthermore, the measured PA of the line-effect stars shows great 
consistency with PA constraints from other techniques, providing further evidence  
that the tool is capable of discovering and constraining circumstellar disks. 

\articlefiguretwo{agcar_tri2.ps}{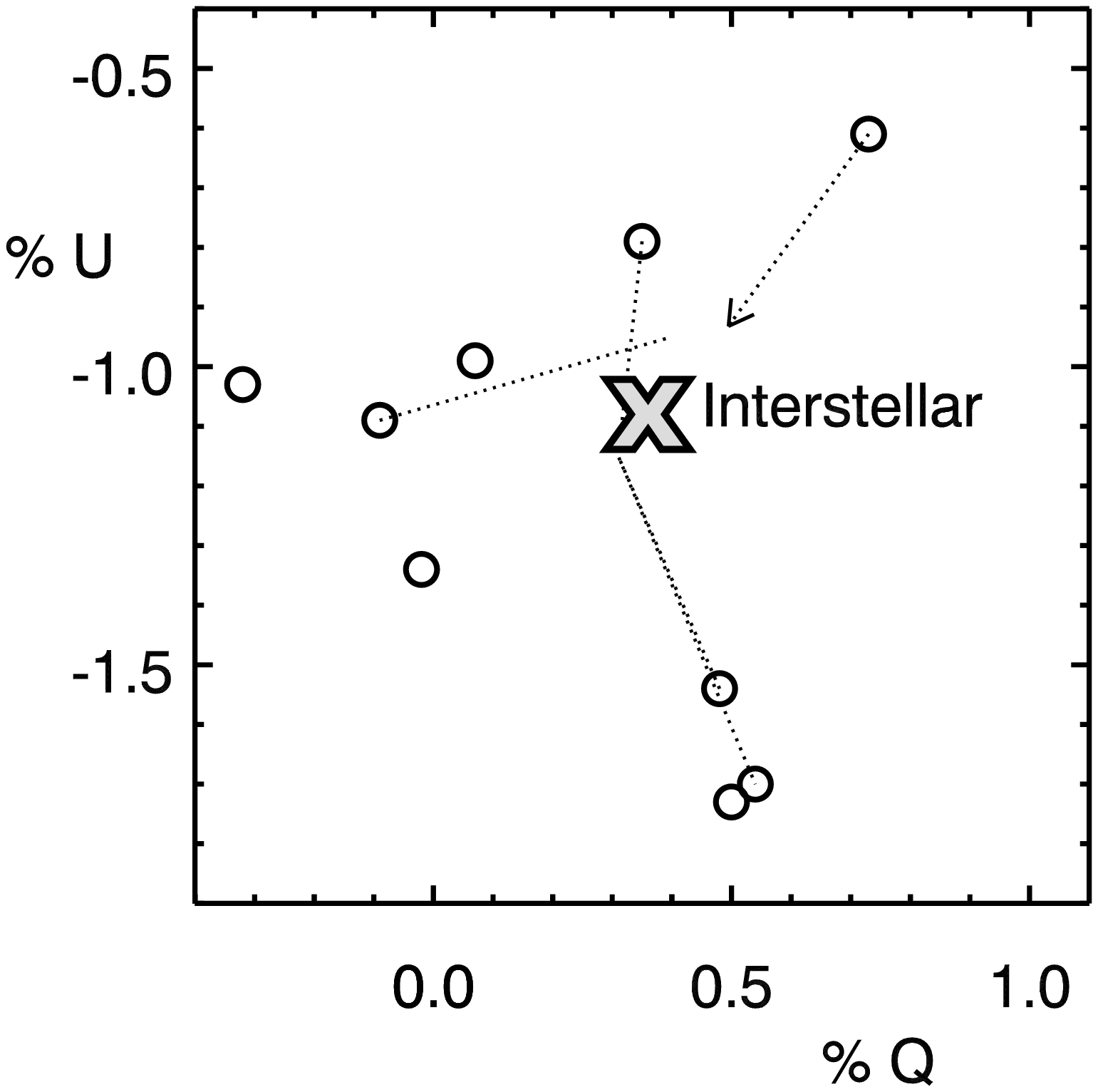}{agcar}{H$\alpha$ polarimetry of the Luminous Blue Variable AG\,Car. 
The triplot on the left hand side reveals line depolarization. 
The large cross on the right hand side denotes the measured polarization at line center 
(constant with time), whilst the open circles represent the continuum measurements 
that vary with time, indicating wind clumping. See Davies et al. (2005, 2007) for details.}

Davies et al. (2005) performed a spectropolarimetry survey of LBVs in the Galaxy and Magellanic Clouds and 
found some intriguing surprises. At first sight the results suggested the presence of disks (or equatorial outflows), 
as the incidence rate of line effects was inferred to be $>$50\%. 
This is notably higher than that of their evolutionary 
neighbors O and Wolf-Rayet stars, with incidence rates of $<$25\% 
(Vink et al. 2009) and $\sim$15\% (Harries et al. 1998) respectively. 
However, when Davies et al. plotted the results of AG\,Car in a $QU$ diagram (see Fig.\,\ref{agcar}) they
noticed that the level of polarization varied with time, which was interpreted as the manifestation
of wind clumping. 

Subsequent modelling by Davies et al. (2007) 
shows how time-variable linear polarization might become a powerful tool to constrain clump sizes and numbers. 
These constraints have already been employed in theoretical studies regarding the origin of wind 
clumping in sub-surface convection layers (Cantiello et al. 2009) and the effects of wind clumping on predicted mass-loss 
rates (Muijres et al. 2011).

One may attempt to derive the number of convective cells by dividing 
the stellar surface area by the size of a convective cell. 
For main-sequence O-type stars, pressure scale heights are in a range
0.04-0.24 $R_{\odot}$, corresponding to a total number of clumps of 
$6 \times 10^3-6 \times 10^4$. These numbers may be tested through
linear polarization variability observations:

\articlefigure[width=8.8cm]{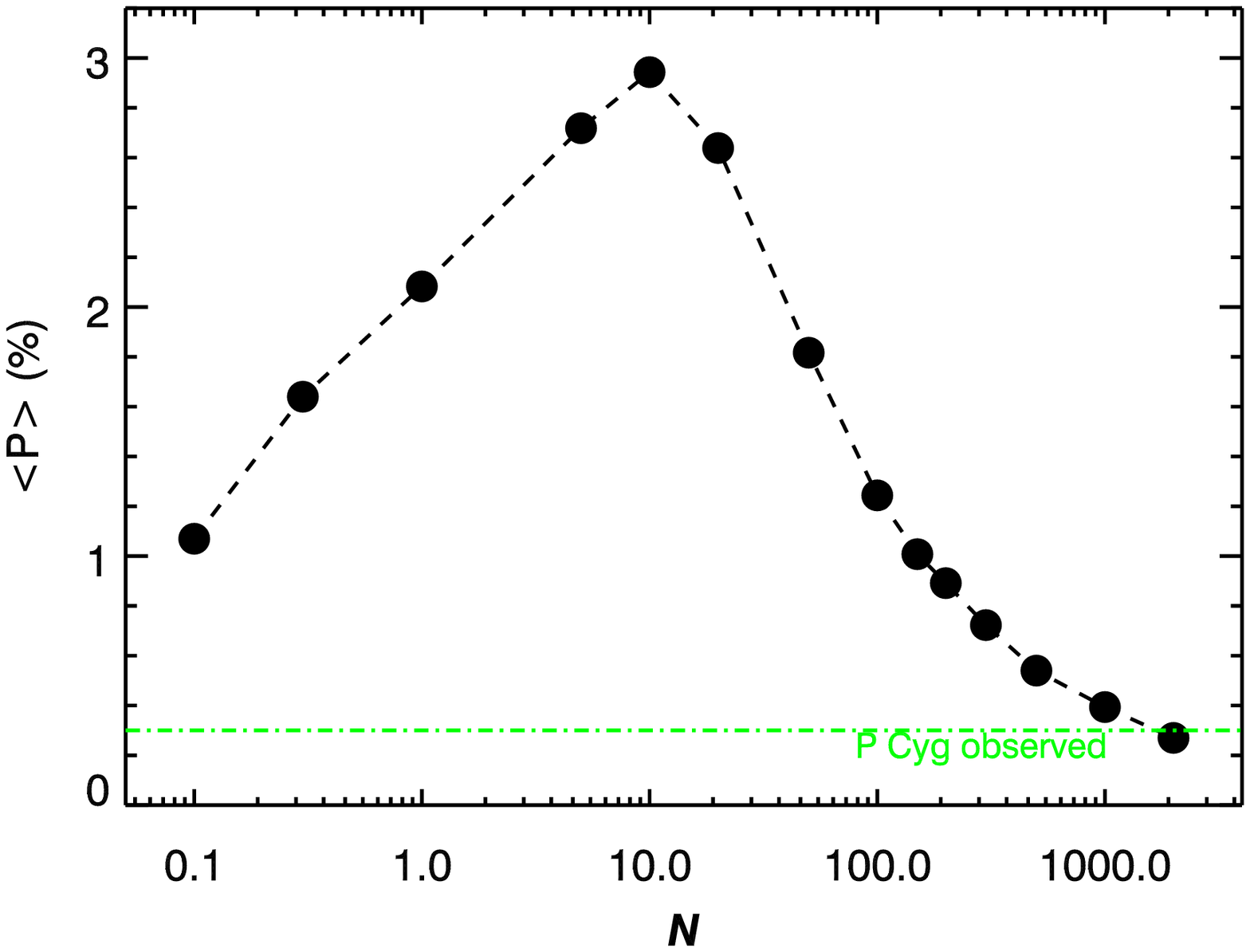}{pcyg}{Time-averaged polarization over a range of ejection rates (per
wind flow-time). At $\mathcal{N} \sim 20$, the optical depth
per clump exceeds unity and the overall polarization drops (see
Davies et al. 2007 for details). The observed amount of polarization 
is indicated by the dash-dotted line. Note that 
there are two ejection-rate regimes where the observed level of polarization 
may be achieved for P\,Cygni.}

Davies et al. (2005) showed that more than half the luminous blue variables (LBVs) are intrinsically polarized. 
As the polarization angle was found to change 
irregularly with time, the observed line effects were considered to be the result of 
wind clumping. 
An example of a model predicting the time-averaged
polarization for the LBV P\,Cygni is shown in Fig.~\ref{pcyg}. 
There are two
regimes where the observed polarization level may be reached.
One is where the ejection rate is very low and only a few very optically thick
clumps are ejected; the other scenario involves that of a very large number of clumps. 
The two scenarios may be
distinguished via time-resolved polarimetry. Given the short
timescale of the polarization variability data, we assume 
that LBV winds consist of thousands of clumps near the photosphere. 
However, for main-sequence O stars the derivation of 
wind-clump sizes from polarimetry 
has not yet been feasible as very high signal-to-noise data 
are required. 
LBVs offer an excellent opportunity for constraining the number of 
clumps from polarization variations because of the 
combination of higher mass-loss rates and lower terminal velocities. 
Davies et al. (2007) found that in order to produce the  
polarization variability of P\,Cygni, the wind 
should consist of some 1000 clumps per wind flow-time. 
In order to check whether this is compatible with subsurface 
convection being the root cause for 
clumping, we would need to consider the sub-surface convective regions 
of an object with properties similar to those of P\,Cygni. 
Because of the lower gravity of P\,Cyg 
the pressure scale height is about 4 $R_{\odot}$, i.e. significantly 
larger than for O-type stars. 
Therefore, the same estimate for the number of clumps as done for the O stars 
yields some 500 clumps per wind-flow time, which is consistent  
with that derived from the polarization data.

\subsection{Monte Carlo disk models}

\articlefiguretwo{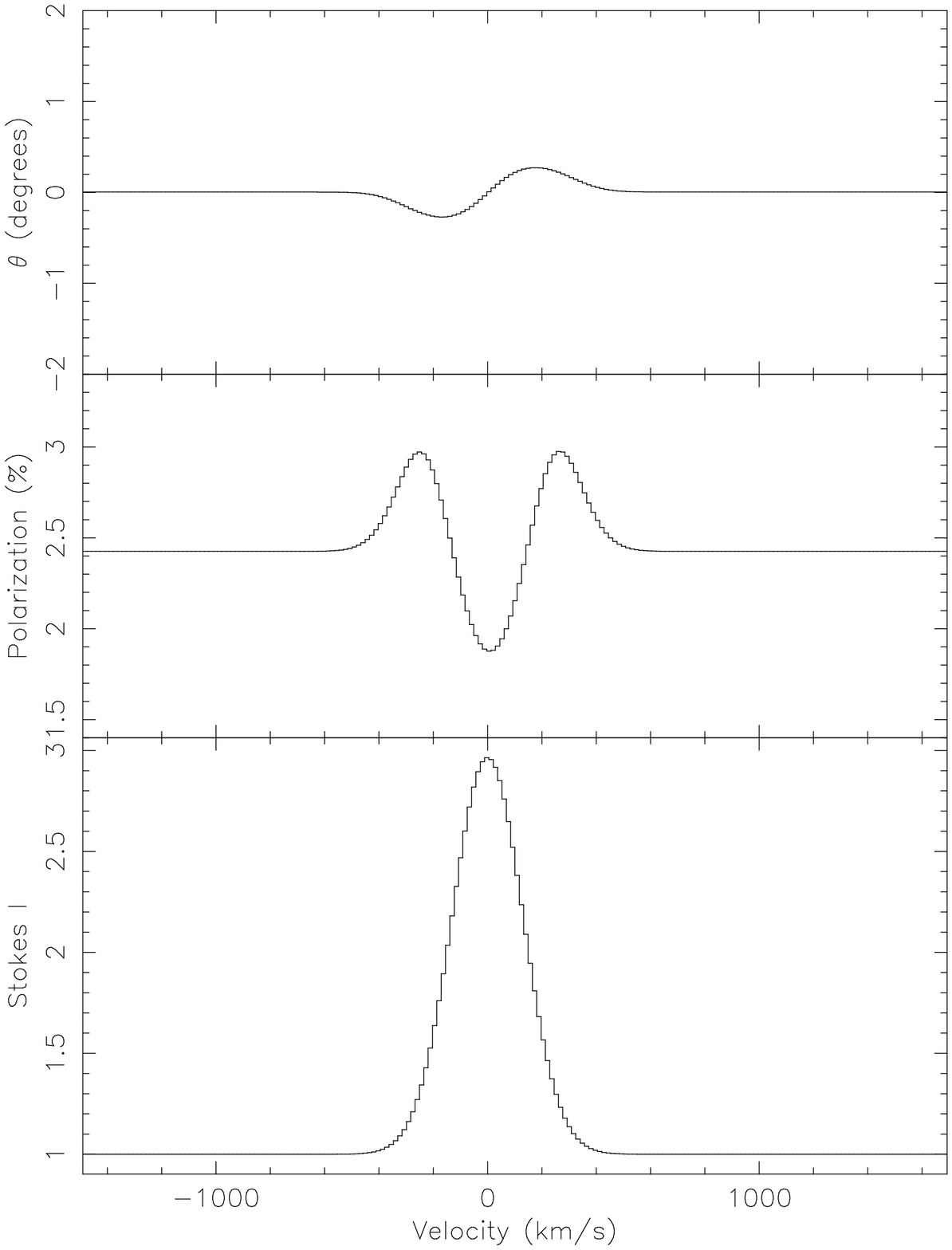}{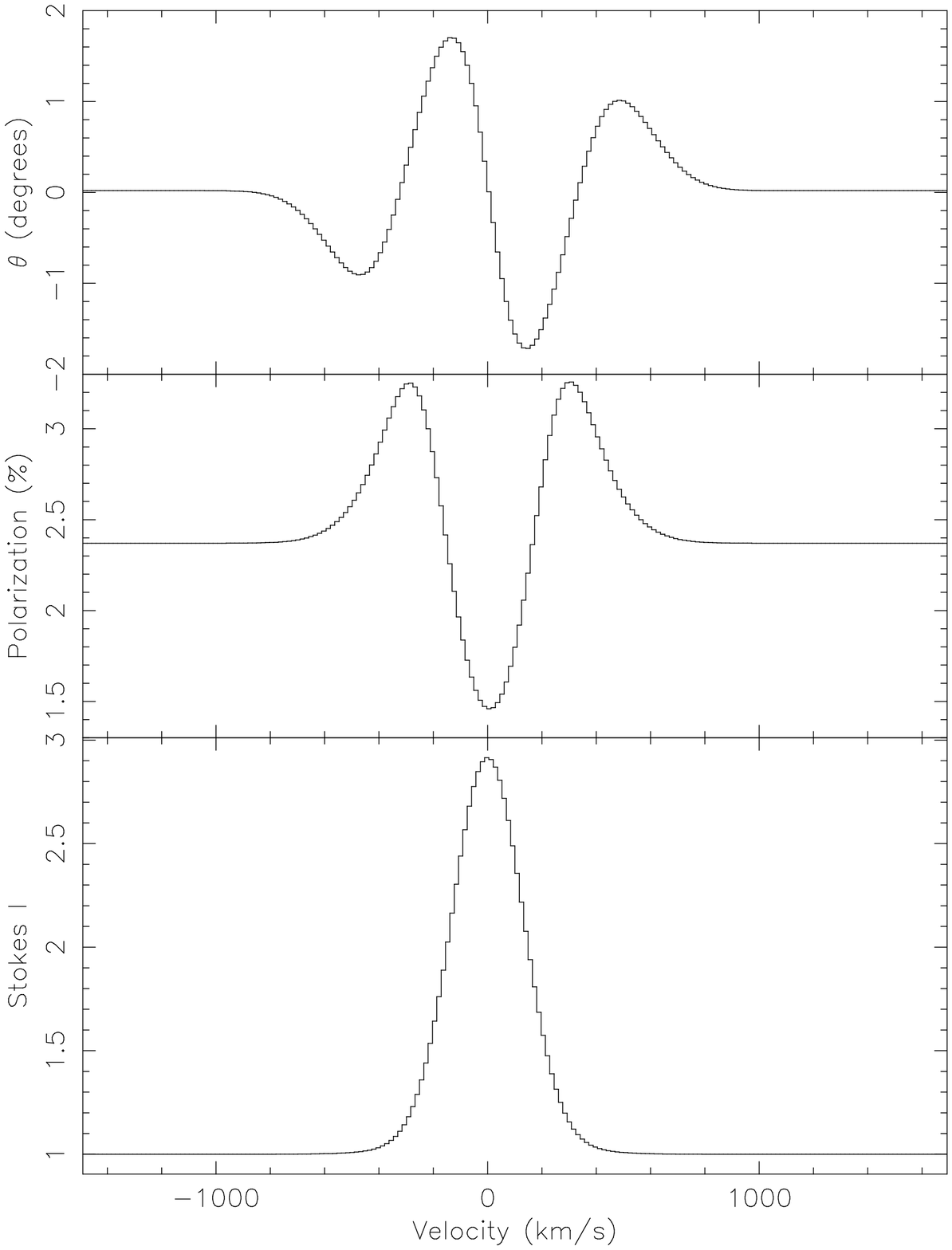}{nohole}{Monte Carlo line polarimetry for the case of a disk with an inner hole (left hand side) 
and without an inner hole -- a result of the finite size of the 
star (right hand side). From Vink, Harries \& Drew (2005a).}

We now return to disks. 
Motivated by the almost {\it ubiquitous} incidence of $QU$ loops in T Tauri and Herbig Ae stars 
(Vink et al. 2002, 2003, 2005b), Vink et al. (2005a)
decided to develop numerical polarization models of line emission scattered off 
Keplerian rotating disks using a 3D Monte Carlo code, with and without a disk inner hole. 
Figure\,\ref{nohole} shows a marked difference between scattering off a disk that reaches the stellar photosphere 
(right hand side), and a disk with a significant inner hole (left hand side). 
The single PA flip on the left-hand side is similar to that predicted analytically (Wood et al. 1993), but 
the double PA flip on the right-hand side -- associated with the undisrupted disk -- came as a surprise at the time. 
The effect is the result of the geometrically correct treatment of the finite-sized stars that interacts with the disk's rotational 
velocity field. 
Our numerical models demonstrated the potential of {\it line} polarimetry (as opposed to depolarization, where 
no velocity information can be gleaned) in establishing not only the disk inclination, but also the size of 
disk inner holes. 

As far as we are aware linear {\it line} polarimetry is as yet the only method 
capable of determining disk holes sizes on the required spatial scales, to test both 
magnetospheric accretion in the young star context, and wind driving for massive stars, as either 
of these processes are crucially determined within the first few stellar radii from the stellar surface.

\subsection{Evidence for disk rotation in B[e] stars}

\articlefigure[width=.9\textwidth]{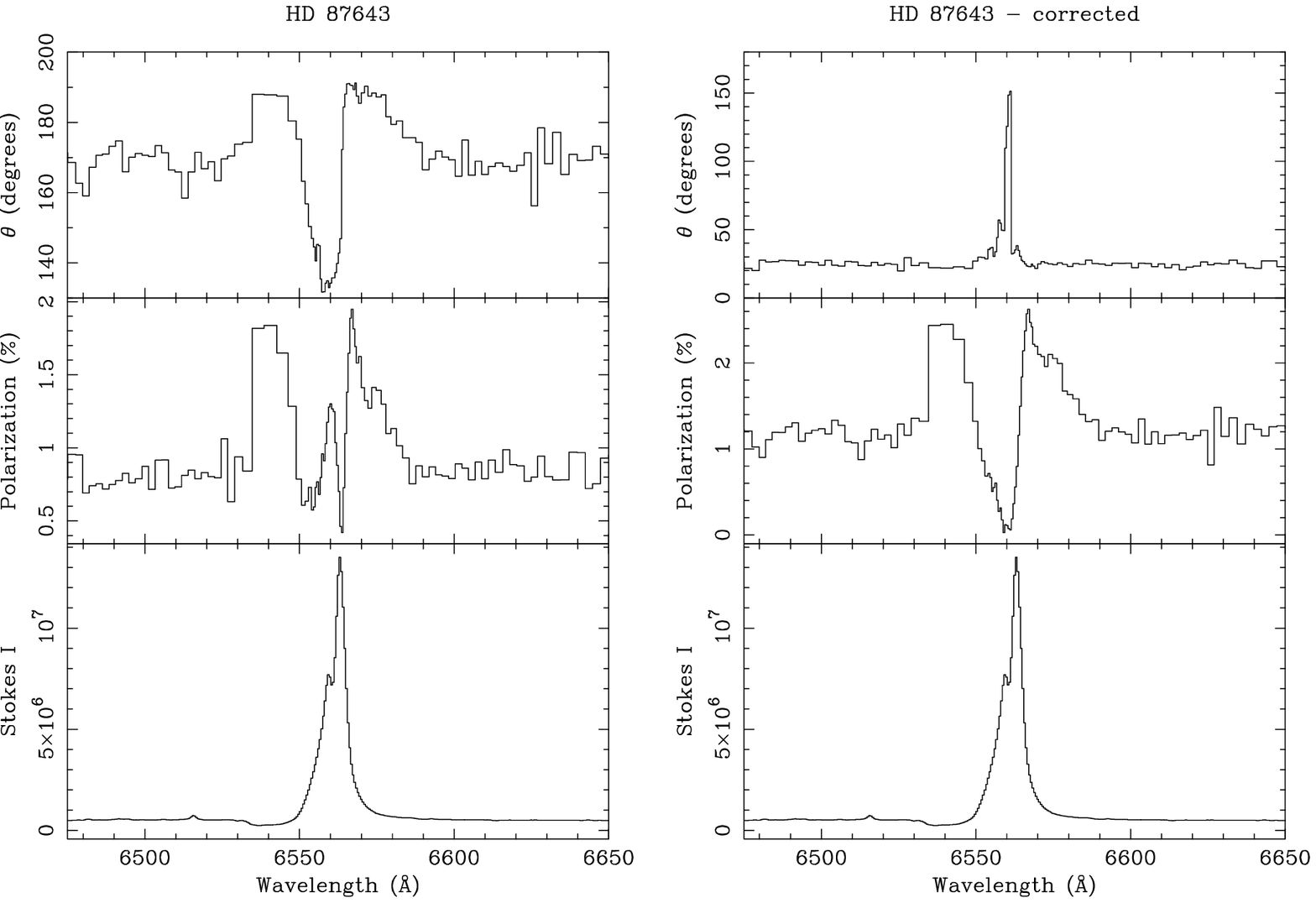}{hd87643}{The polarization spectrum of the B[e] star 
HD 87643, and the corrected spectrum of the star. The PA rotation suggests rotation of the disk. 
For more details, see Oudmaijer et al. (1998) and Oudmaijer \& Drew (1999).}

Returning to evolved objects such as B[e] stars we may perform similar observations as 
we performed on PMS stars. 
The best example of line spectropolarimetry in B[e] stars is given in Fig.\,\ref{hd87643}.
It shows the unambiguous presence of rotation as noted in the change of the position angle.
Spectroscopic and spectropolarimetric data for this object have
been presented and analyzed in detail by Oudmaijer et al. (1998) and Oudmaijer \& Drew (1999). 
A comparison with the schematic model calculations by Wood et al. (1993) and Vink et al. (2005a) 
indicates that the polarization profile may indeed be reproduced with a rotating circumstellar disk.

\section{Outlook}

The first question to address is whether progress is expected to 
mostly come from theory, modelling or observations. Monte Carlo models 
(e.g. for disks with and without inner holes as shown in Fig.\,\ref{nohole}) are rather 
versatile, so modelling is not an issue. However, in order to be able to 
test the subtle differences shown in in Fig.\,\ref{nohole}, we require data with higher S/N.

We are at an exciting time in history, as we are on the verge of making a huge breakthrough. 
Normal stellar Stokes I spectroscopy at sufficient S/N has been performed for over 100 years, but because linear
spectropolarimetry is photon hungry (as for most relevant cases the degree of polarization is just of 
order of $\sim$1\%), we are about to enter a completely new era: that of of extremely 
large telescopes, providing us sufficient S/N to utilize the tool to its full potential.

A second new aspect will be time dependence. For B[e] stars linear polarization measurements have 
generally been attributed to the presence of disks (Magalhaes 1992; Oudmaijer \& Drew 1999), 
but before we can unquestionably proof this, we 
would need to check that the PA is constant with time. The situation of the LBVs, where we now know the 
PA is variable (Davies et al. 2005), should be considered as a warning sign.

Finally, spectropolarimetric monitoring will be needed in order to 
probe 3D structures, such as wind clumping and/or rotating disk structures. 
This may become possible with future space instruments such as Arago/UVMag (Neiner 2015).  



\end{document}